\documentclass[12pt]{article}
\pdfoutput=1
\usepackage[top=30truemm,bottom=30truemm,left=25truemm,right=25truemm]{geometry}
\usepackage{authblk}
\usepackage{amsthm}
\usepackage{amsmath,amssymb,bm}
\usepackage{cite}
\usepackage[]{hyperref}
\hypersetup{colorlinks,linkcolor={blue},citecolor={blue},urlcolor={blue}}  
\usepackage[titletoc]{appendix}
\usepackage{url}
\usepackage{graphicx,color}
\usepackage{braket}
\usepackage{qcircuit}
\usepackage{here}
\usepackage{mathrsfs}

\usepackage{titlesec}

\titleformat*{\section}{\large\bfseries}

\newtheorem*{definition*}{Definition}
\newtheorem*{assumption*}{Assumption}

\def\sinh{{\mathrm{sinh}}}
\def\cosh{{\mathrm{cosh}}}
\def\tanh{{\mathrm{tanh}}}

 \def\CM{{\cal M}}

\def\be{\begin{equation}}
\def\ee{\end{equation}}
\def\ba{\begin{eqnarray}}
\def\ea{\end{eqnarray}}

\AtBeginDocument{%
  {\footnotesize\global\footnotesep=0.8\baselineskip}%
}

\begin{document}

\begin{titlepage}
\thispagestyle{empty}

\begin{flushright}
\end{flushright}

\bigskip

\begin{center}
\noindent{\bf \large Holographic Dual of Crosscap Conformal Field Theory
}\\
\vspace{1.4cm}

{\bf Zixia Wei}
\vspace{1cm}\\

{\it
Jefferson Physical Laboratory, 
Harvard University, Cambridge, MA 02138, USA
}\\[1.5mm]


\vskip 3em
\end{center}

\begin{abstract}
We propose a holographic dual for 2D CFT defined on closed non-orientable manifolds, such as the real projective plane $\mathbb{RP}^2$ and the Klein bottle $\mathbb{K}^2$. Such CFT can be constructed by introducing antipodally identified cuttings, i.e. crosscaps, to a sphere and hence called crosscap CFT (XCFT). 
The gravity dual is AdS$_3$ spacetime with dS$_2$ end-of-the-world branes. 
In particular, the Lorentzian spacetime with a global dS$_2$ brane is dual to the unitary time evolution of a crosscap state in CFT, post-selected on the CFT ground state. 
We compute the holographic $\mathbb{RP}^2$ partition function (or the $p$-function), one-point function, and $\mathbb{K}^2$ partition function, and see that they successfully reproduce the XCFT results. We also show a holographic $p$-theorem as an application. 
\end{abstract}

\end{titlepage}

\newpage
\setcounter{page}{1}
\tableofcontents


\section{Introduction}
As the most well-studied example of the holographic principle \cite{tHooft93,Susskind94}, the AdS/CFT correspondence has played a central role in our understanding of AdS quantum gravity in a nonperturbative way \cite{Maldacena97}.  
In this paper, we construct a minimal {bottom-up} model for describing the AdS$_3$ dual of 2D CFT defined on non-orientable manifolds and find that a structure of dS$_2$ vacuum naturally appears in this construction. Being a standard holographic CFT, the boundary theory is unitary and local and has a standard Hilbert space and a canonical time evolution.

\begin{figure}
    \centering
    \includegraphics[width=12cm]{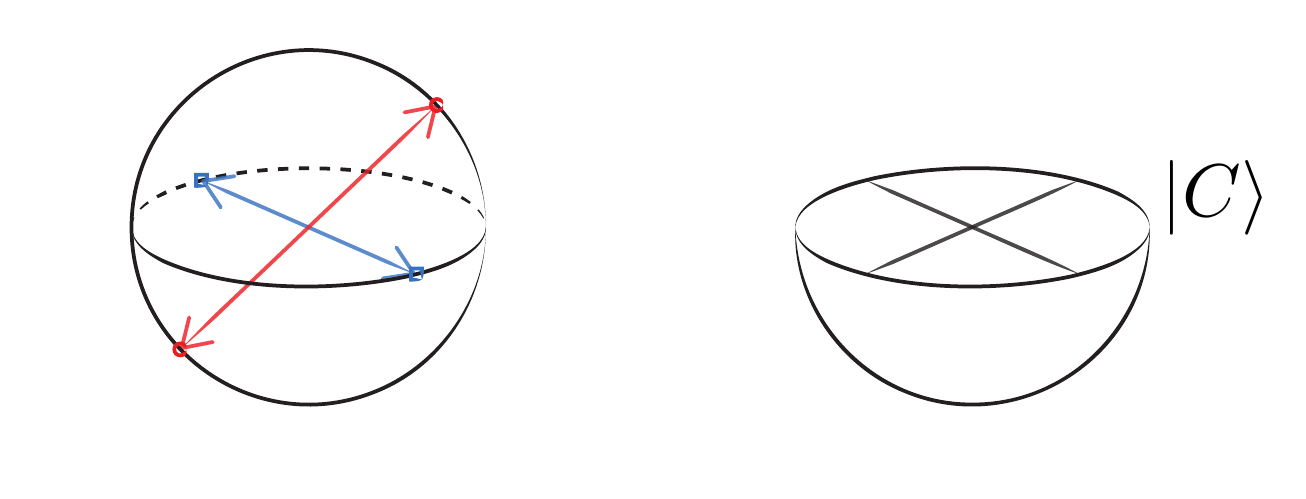}
    \caption{The real projective plane $\mathbb{RP}^2$ is defined by imposing an antipodal identification to a sphere $S^2$. This is equivalent to identifying the antipodal points on the boundary of a hemisphere, i.e. inserting a crosscap denoted as $\otimes$. The state introduced by the crosscap is denoted as $\ket{C}$.}
    \label{fig:RP2_crosscap}
\end{figure}

The most simple 2D non-orientable manifold is the real projective plane $\mathbb{RP}^2$, which is defined by identifying the antipodal points on a sphere $S^2$. Topologically, this is equivalent to starting from a hemisphere or a disk and then identifying the antipodal points on the boundary as sketched in Fig.\ref{fig:RP2_crosscap}. The structure obtained by identifying the antipodal points on the boundary is called a crosscap.  
Since any 2D non-orientable manifold can be constructed by introducing crosscaps to an orientable manifold, we may call the CFT defined on such manifolds the crosscap CFT (XCFT). Historically, XCFT has been extensively studied in the context of string world sheet theory with orientifolds \cite{FPS93,PSS95,BP09}. We will be mainly focusing on holographic XCFT defined on $\mathbb{RP}^2$ and the Klein bottle $\mathbb{K}^2$, which are topologically equivalent to inserting one and two crosscaps into a sphere, respectively.

Looking for the gravity dual of a 2D XCFT at the semi-classical level presents an apparent obstacle. As is known in cobordism theory, an $\mathbb{RP}^2$ does not bound any compact 3D manifold\footnote{Note that we do not need to require the 3D manifold to be a smooth manifold \cite{KS77,MM79}.
The original cobordism theory developed by Thom was for smooth manifolds, i.e. manifolds with smooth structures. However, one may also consider other versions of cobordism theory for manifolds with more or less structures. The cobordism theory for topological manifolds (i.e. manifolds with no additional structures), called the topological cobordism theory, was established by Kirby and Siebenmann \cite{KS77}. The topological cobordism theory is of course not equivalent to the original cobordism theory, but they turn out to be equivalent in 2D. See the table on page 324 in \cite{KS77}. This is because any 3D and lower dimensional topological manifold admits a unique smooth structure due to the Moise's theorem. See page 308 in \cite{KS77}.
Also, note that this statement implies that there does not exist a 3D asymptotically AdS manifold whose 
conformal compactification is compact and has no boundary other than the $\mathbb{RP}^2$ asymptotic boundary.
}. Therefore, new elements beyond pure gravity need to be introduced to resolve this issue.
In our construction of holographic duals of XCFT$_2$, a dynamical end-of-the-world brane which is locally dS$_2$ is introduced and plays the most central role throughout the results presented in this paper. 
This makes a major distinction from previous discussions on the holographic dual of crosscaps and non-orientable CFTs including \cite{MNSTW15,Verlinde15,NO15,MR16,GMT16,NO16,LTV16,GKZ20,Tsiares20,CK21,CR22}. 

In the following, we start by presenting the holographic dual of XCFT defined on $\mathbb{RP}^2$ and writing down the action of our minimal {bottom-up} construction. We will compute the holographic partition function and 1-point functions and see their matching with standard $\mathbb{RP}^2$ XCFT results.
We then move on to the Lorentzian counterpart and explain how a dS$_2$ vacuum is described by a conformal crosscap state under unitary time evolution and post-selected on the CFT ground state. We will also point out an intriguing relation between the conformal crosscap state and the Type II$_1$ von Neumann algebra.  
As a slightly more complicated example, we will analyze the holographic dual of $\mathbb{K}^2$ and show that the holographic partition function matches known universal results in the XCFT and shows a first order phase transition analogous to the Hawking-Page transition.
We will end up with an application of our AdS/XCFT correspondence by showing a holographic $p$-theorem, which states that there exists a holographic $p$-function which monotonically decreases under a holographic renormalization group (RG) flow, {whose XCFT counterpart has not been found yet.}

\section{General Construction and Gravity Dual of \texorpdfstring{$\mathbb{RP}^2$}{RP2} XCFT}
Let us start by considering a holographic CFT\footnote{Throughout this paper, we use the terminology ``holographic CFT" to refer to a CFT which admits a semiclassical gravity dual described by the Einstein gravity, without specifying its characterization in terms of the CFT data such as the sparse light spectrum condition \cite{HPPS09, SvRW20}. It would be an interesting future direction to characterize the OPE data for the XCFT which admits a gravity dual described in this paper.} on a unit sphere $S^2$ parameterized by $\theta\in[-\frac{\pi}{2}, \frac{\pi}{2}]$ and $\phi\in[0,2\pi)$ with $\phi\sim\phi+2\pi$. The gravity dual is given by the Euclidean AdS$_3$
\begin{align}\label{eq:global_metric}
    ds^2 = d\eta^2 + \sinh^2 \eta \left(d\theta^2 + \cos^2 \theta d\phi^2 \right). 
\end{align}
For simplicity, we set the AdS radius to $1$ throughout this paper. 
Let us then impose the antipodal identification $(\theta,\phi)\sim(-\theta,\phi+\pi)$ to squash the boundary manifold $S^2$ into an $\mathbb{RP}^2$. If we were to directly extend this identification into the bulk, a fixed point would appear at $\eta=0$ and would turn out to be a singularity. 
{Although this bulk geometry can recover certain features of XCFT on $\mathbb{RP}^2$ \cite{MR16,NO16,GKZ20}, it is globally not a smooth solution of the bulk Einstein equation and the variational principle at the neighborhood of the singularity is not understood. As a result, whether it contributes to the gravitational path integral or not, and how much it contributes if the answer to the previous question is positive, cannot be determined straightforwardly using the standard dictionary of AdS/CFT  \cite{GKP98,Witten98,Witten07,MW07}.}
{Note that this singularity cannot be resolved by engineering the 3D action or allowing complex metrics, since an $\mathbb{RP}^2$ cannot be the boundary of any (smooth) 3D manifold, according to the standard results in cobordism theory.}

In order to resolve this singularity, we propose a minimal model by introducing an end-of-the-world brane $Q$ in the bulk $\mathcal{M}$. 
For a holographic CFT defined on a closed manifold $\Sigma$, we consider the dual bulk (Euclidean) action given by 
\begin{align}\label{eq:action}
    I = -\frac{1}{16\pi G_N} \int_{\CM} \sqrt{g} \left(R-2\Lambda\right) - \frac{1}{8\pi G_N} 
\int_{Q} \sqrt{h}(K-T) - \frac{1}{8\pi G_N} 
\int_{\Sigma} \sqrt{\gamma}B, 
\end{align}
where the three terms are the Einstein-Hilbert term in $\mathcal{M}$, the Gibbons-Hawking term on the brane $Q$, and the Gibbons-Hawking term on the asymptotic boundary $\Sigma$, respectively. 
$G_N$, $g_{\mu\nu}$, $R$ {and $\Lambda = -1$} are the Newton constant, the metric, the scalar curvature and the cosmological constant in $\mathcal{M}$. $h_{ab}$, $K_{ab}$, $T$ are the induced metric, the extrinsic curvature and the tension of $Q$. $\gamma_{ij}$ and $B_{ij}$ are the induced metric and the extrinsic curvature on the asymptotic boundary $\Sigma$. Note that we have defined the extrinsic curvatures of $\Sigma$ and $Q$ such that the normal vectors are pointing outwards from the bulk $\mathcal{M}$.
While imposing the standard Dirichlet boundary condition on $\Sigma$, we choose the Newmann boundary condition 
\begin{align}
    K_{ab} - K h_{ab} = -T h_{ab},
\end{align}
on the brane $Q$ to make it dynamical. In order to get a sensible solution in later discussions, we restrict the tension to $T<-1$.  

\begin{figure}
    \centering
    \includegraphics[width=16cm]{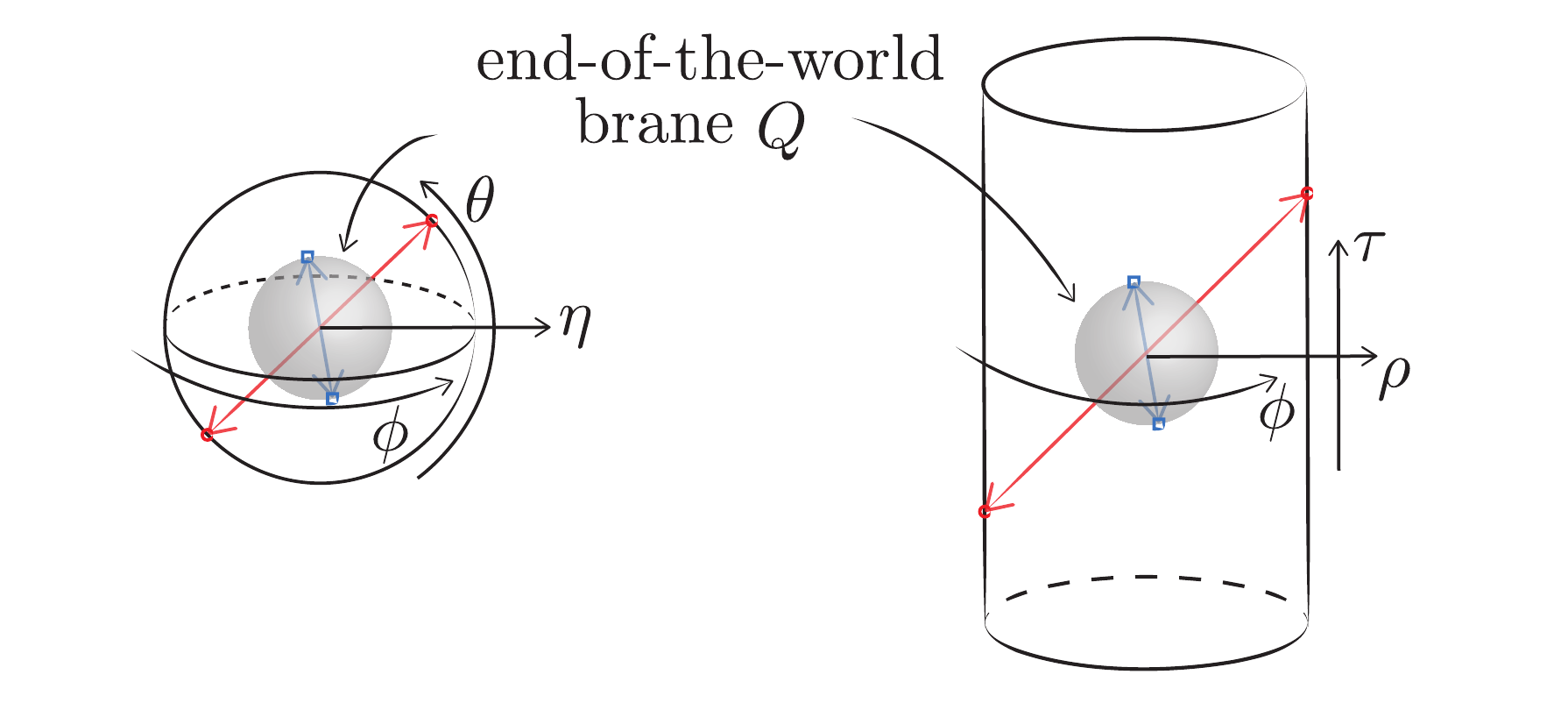}
    \caption{(Left) The holographic dual of XCFT on $\mathbb{RP}^2$ is given by the $\eta\geq\eta_*$ region of the $\mathbb{Z}_2$ quotient $(\theta,\phi)\sim(-\theta,\phi+\pi)$ of AdS$_3$ with an $\mathbb{RP}^2$ (or, equivalently, Euclidean dS$_2/\mathbb{Z}_2$) end-of-the-world brane $Q$ at $\eta=\eta_{*}$. If $Q$ were not introduced, there would be a problematic singularity at $\eta = 0$. (Right) The $(\rho,\tau,\phi)$-coordinate is useful for considering holographic XCFT defined on a half-cylinder with a crosscap.}
    \label{fig:RP2_holo_dual}
\end{figure}

Solving the Einstein equation in $\mathcal{M}$ with the asymptotic boundary given by $\mathbb{RP}^2$, we get a $\mathbb{Z}_2$ quotient of global AdS$_3$ obtained by $(\theta,\phi)\sim(-\theta,\phi+\pi)$, with an end-of-the-world brane $Q$ located at 
\begin{align}\label{eq:etastar}
    \eta = {\rm arccoth}(-T) \equiv \eta_*. 
\end{align}
The {intrinsic} geometry of $Q$ is {$\mathbb{RP}^2$}. The bulk $\mathcal{M}$ is defined at $\eta\geq\eta_*$ and the $\eta<\eta_*$ region is excluded by introducing the brane $Q$. Therefore, there is no singularity in the bulk any more. {In terms of cobordism, after introducing the end-of-the-world brane, the whole boundary of the bulk becomes $\mathbb{RP}^2 \cup \mathbb{RP}^2$, which can bound a 3D manifold.}\footnote{While we are merely discussing bottom-up constructions in 3D gravity, note that introducing the end-of-the-world brane is also natural from the top-down point of view. In top-down constructions via string theory (or M-theory), 3D gravity should arise from the dimensional reduction of 10D (or 11D) gravity, which is asymptotically ${\rm AdS}_3 \times X$, where $X$ is a compact manifold of 7D (or 8D). When the corresponding CFT is defined on $\mathbb{RP}^2$, we are looking for a 10D (or 11D) bulk geometry whose asymptotic boundary is $\mathbb{RP}^2 \times X$, assuming a geometric dual exists. In this scenario, since $\mathbb{RP}^2$ can never cap off in the bulk, it must be the $X$ direction that smoothly caps off. After performing the dimensional reduction to 3D, this smooth cap-off will appear as an end-of-the-world brane.}
See Fig.\ref{fig:RP2_holo_dual} for a sketch.

Let us check how the symmetries match between the XCFT side and the gravity side in terms of the global conformal group. On the XCFT side, realizing the $\mathbb{RP}^2$ by identifying the antipodal points on the Riemann sphere $S^2$, it is easy to see that the isometry group of $\mathbb{RP}^2$ is $PO(3) = PSO(3)\cong SO(3)$. This $SO(3)$ is inherited from the $SO(3)$ isometry of $S^2$, which is a subgroup of its conformal group $SO(3,1)$. On the other hand, the other half of the conformal symmetries are broken, since they do not map great circles to great circles on $S^2$ and hence break antipodal relations. This $SO(3)$ symmetry manifests in the Euclidean gravity dual. The existence of the $\mathbb{RP}^2$ (or, equivalently, Euclidean dS$_2/\mathbb{Z}_2$) end-of-the-world brane, which essentially introduce a center in the bulk spacetime, breaks the three translational symmetries but preserve three rotational symmetries, which form the $SO(3)$. 

It is useful to introduce another coordinate system by performing 
\begin{align}
    \cos \theta = \frac{\sinh \rho}{\sinh \eta}, ~~ \sin \theta = \frac{\tanh \tau}{\tanh \eta}. 
\end{align}
In the new coordinate system, the metric is  
\begin{align}\label{eq:Emetric_global}
    ds^2 = d\rho^2 + \cosh^2 \rho~ d\tau^2 + \sinh^2 \rho ~d\phi^2,
\end{align}
the identification is $(\tau,\phi) \sim (-\tau,\phi+\pi)$, and the brane profile is 
\begin{align}\label{eq:Ebrane_global}
    T^2 = \frac{\cosh^2 \rho}{\sinh^2 \rho + \tanh^2 \tau},
\end{align}
as shown in Fig.\ref{fig:RP2_holo_dual}. If we introduce a cutoff at $\rho = \rho_\infty$, then the corresponding XCFT is defined on a half-infinite cylinder with a crosscap inserted at $\tau=0$. 

{Before proceeding, we would like to note that there are two possible ways to interpret the brane term introduced in the action Eq.\eqref{eq:action}. One is to introduce the brane only when there are crosscaps on the CFT side. This way of thinking is similar to that appears in the construction of the AdS dual corresponding to a CFT defined on a manifold with boundaries (called the boundary CFT (BCFT)) \cite{Takayanagi11,FTT11}, where AdS end-of-the-world branes are introduced only when there exist boundaries on the CFT side. Another way of thinking is to include the brane term regardless of whether there are crosscaps on the CFT side or not, and allow all the solutions in the gravitational path integral including brane nucleations. Both ways of thinking give the same results as long as we only look at the most dominating bulk configurations. However, it would be an interesting future question to discuss their differences when subdominating contributions are included. Especially, the second way of thinking may mediate a Coleman-de Luccia instability \cite{CdL80} in the bulk.}

\section{Partition Function, Crosscap Entropy and One-point Function}
The partition function of an XCFT defined on a unit $\mathbb{RP}^2$ can be written as the inner product between the CFT ground state $\ket{0}$ and a crosscap state $\ket{C}$ 
\begin{align}\label{eq:ZRP}
    Z_{\mathbb{RP}^2} = \braket{0|C},
\end{align}
and is sometimes called the $p$-function.\footnote{Here we adopt the terminology introduced in \cite{CK21} where $p$ stands for ``parity". } One natural way to see this is to consider the fundamental domain $\theta\in[-\frac{\pi}{2},0]$ of the $\mathbb{RP}^2$. Then the path integral over $\theta\in[-\frac{\pi}{2},0)$ gives the CFT ground state $\ket{0}$ and the crosscap state $\ket{C}$ is the ``boundary" condition imposed on $\theta=0$. (Refer to Fig.\ref{fig:RP2_crosscap}.)
In order for the energy-momentum tensor to match between identified antipodal points, $\ket{C}$ must satisfy the following crosscap condition \cite{Ishibashi88},
\begin{align}\label{eq:Xstate_condition}
    \left(L_n - (-1)^{n} \bar{L}_{-n}\right) \ket{C} = 0, \qquad n\in \mathbb{Z},
\end{align}
where $L_n$'s ($\bar{L}_n$'s) are the chiral (anti-chiral) Virasoro generators. Note that the crosscap condition Eq.\eqref{eq:Xstate_condition} preserves exactly a half of the Virasoro symmetry. 
Besides, the crosscap entropy\footnote{The crosscap entropy is also called the Klein bottle entropy in some literature \cite{Tu17} since it appears in a limit of the Klein bottle partition function. Here we adopt the more general name crosscap entropy.} defined as 
\begin{align}
    S_{\rm xc} = \log\langle0|C\rangle, 
\end{align}
is a quantity closely related to the $p$-function. The $p$-function and the crosscap entropy in XCFT are analogous to the $g$-function and the boundary entropy in BCFT \cite{AL91}, respectively. 

In our AdS/XCFT for $\mathbb{RP}^2$, we have  
\begin{align}
    &R=-6,~~\Lambda = -1,~~\sqrt{g} = \sinh^2 \eta |\cos \theta|, \\
    &K = -2 \frac{\cosh \eta_*}{\sinh \eta_*}, ~~T = - \frac{\cosh \eta_*}{\sinh \eta_*},~~\sqrt{h} = \sinh^2 \eta_* |\cos \theta|. 
\end{align}
Plugging these into the action Eq.\eqref{eq:action}, integrating over the bulk dual and subtracting divergent terms, 
we obtain the holographic $p$-function 
\begin{align}\label{eq:pfunction}
    p \equiv Z_{\mathbb{RP}^2} = \exp\left(-I^{\rm (ren)}\right) = \exp\left( -\frac{\eta_*}{4G_N}\right) = \exp\left( -\frac{c}{6}{\rm arccoth}(-T) \right),
\end{align}
where $I^{\rm (ren)}$ is the renormalized bulk action and $c$ is the central charge of the XCFT. Note that we have used Eq.\eqref{eq:etastar} and the Brown-Henneaux relation $c = \frac{3}{2G_N}$ \cite{BH86} in the last line.
Accordingly, the crosscap entropy is
\begin{align}\label{eq:Sxc}
    S_{\rm xc} = \log |p| = -\frac{\eta_*}{4G_N} = -\frac{{\rm arccoth}(-T)}{4G_N} = -\frac{c}{6}{\rm arccoth}(-T) .
\end{align}
Given the similarity between the $p$-function in XCFT and the $g$-function in BCFT, we may expect that there is a ``$p$-theorem" stating that the $p$-function monotonically decreases under some RG flow. While how to formulate such a ``$p$-theorem" is not yet known in XCFT, we will later come back to the holographic $p$-function and show a holographic $p$-theorem. 

It is also straightforward to check that holographic 1-point functions for scalar primaries match the XCFT result. Consider an XCFT defined on  $\mathbb{RP}^2$ with radius $r$ and insert a scalar primary $\mathcal{O}$ with scaling dimension $\Delta$ at $(\theta,\phi)$. The 1-point function is determined by the symmetries except for the coefficient, and has the form 
\begin{align}\label{eq:1point}
    \langle \mathcal{O}(\theta,\phi) \rangle_{\mathbb{RP}^2} = \frac{a_{\mathcal{O}}}{(2r)^{\Delta}}.
\end{align}
On the gravity side, the most simple regime is given by {the case when $\Delta$ is $O(c^0)$} but at the same time large $\Delta\gg1$. In this regime, the primary operator corresponds to a free scalar field with mass $m=\sqrt{\Delta(\Delta-2)}$. With the presence of the brane $Q$, the holographic 1-point function can be approximated by $\lambda e^{-\Delta L}$, where $L$ is the distance between the boundary operator insertion and the brane $Q$ (with proper renormalization performed) and $\lambda$ is a constant depending on the details of the boundary condition imposed on the brane $Q$ for the scalar field \cite{FTT11,KS21}. Since we are considering a holographic CFT defined on $\mathbb{RP}^2$ with radius $r$, we can regard our XCFT defined at $\eta = \log(2r/\epsilon)$ where $\epsilon$ is a UV cutoff corresponding to the lattice distance. The distance between $(\log(2r/\epsilon),\theta,\phi)$ and the brane $Q$, after subtracting the divergent term, 
is $L = \log 2r - \eta_*$. Accordingly, the holographic 1-point function is given by $\lambda e^{\eta_*\Delta}/(2r)^{\Delta}$, which has the same form as the XCFT result Eq.\eqref{eq:1point}. It is straightforward (though a little bit more complicated) to check the holographic 1-point functions for scalar primaries with $\Delta\sim1$ and $\Delta=O(c)$ also have the form Eq.\eqref{eq:1point}, {following the same strategy as in \cite{FTT11,KS21,KW22}.}

\section{Lorentzian Signature from Wick Rotation}\label{sec:Lorentzian}

\begin{figure}
    \centering
    \includegraphics[width=14cm]{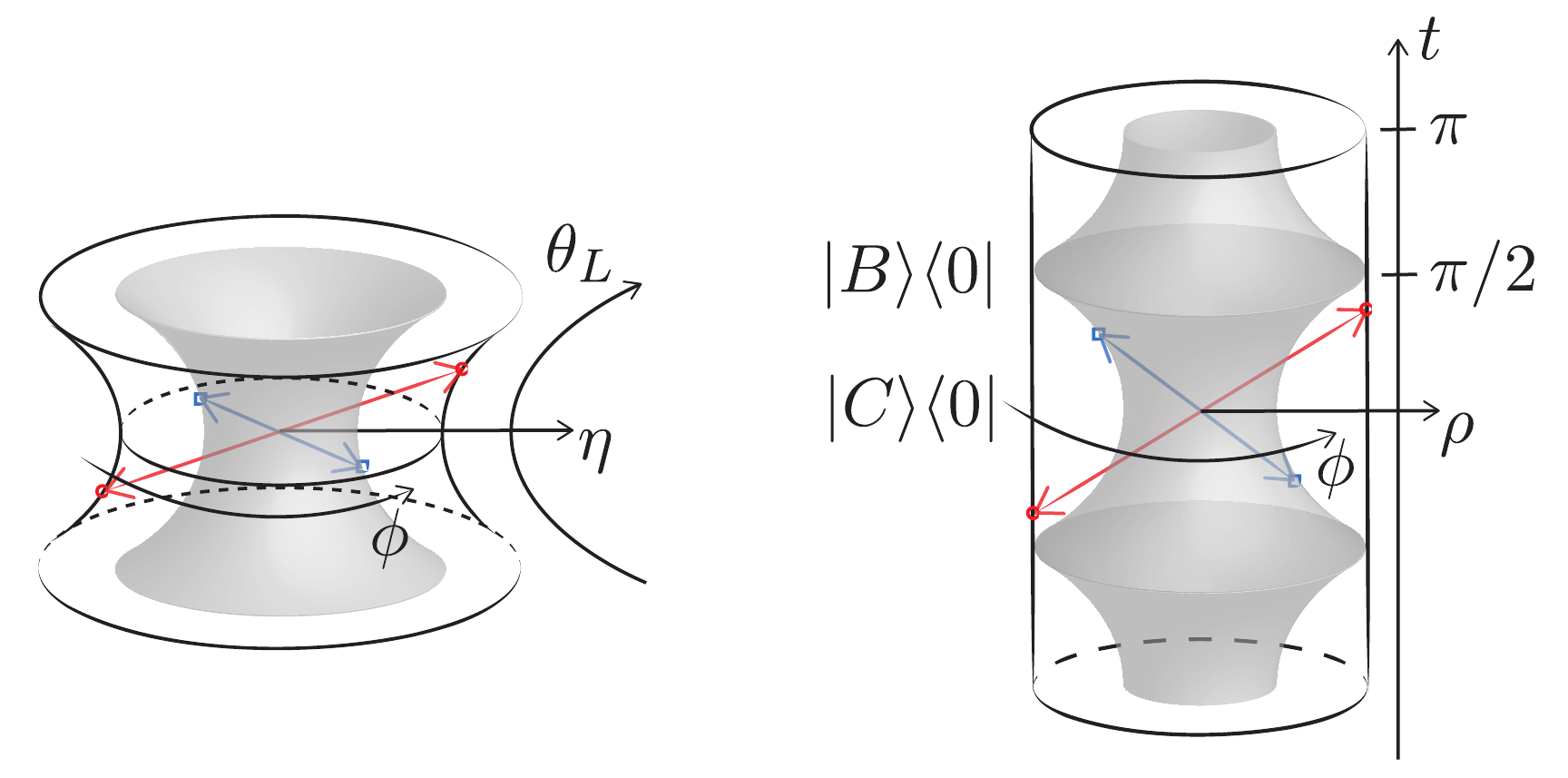}
    \caption{The analytic continuations to the Lorentzian signature of the configurations in Fig.\ref{fig:RP2_holo_dual}. The end-of-the-world brane $Q$ hits the asymptotic boundary at $t=\pi/2$, reflecting the features of the boundary state $\ket{B} = e^{-i\pi H/2}\ket{C}$.}
    \label{fig:RP2_holo_dual_Lorentzian}
\end{figure}

Let us then consider the Wick rotation to the Lorentzian signature in the setups discussed above. First, if we take $\theta \rightarrow i\theta_L$ in 
Eq.\eqref{eq:global_metric}, the metric turns into 
\begin{align}\label{eq:metric_WR}
    ds^2 = d\eta^2 + \sinh^2 \eta \left(-d\theta_L^2 + \cosh^2 \theta_L ~d\phi^2 \right),  
\end{align}
with identification $(\theta_L,\phi)\sim(-\theta_L, \phi+\pi)$. 
The brane $Q$ is still at $\eta=\eta_*$. The bulk geometry is manifestly a warped product between dS$_2/\mathbb{Z}_2$ and a half infinite line $\eta\geq\eta_*$. The boundary geometry is also dS$_2/\mathbb{Z}_2$. See the left panel of Fig.\ref{fig:RP2_holo_dual_Lorentzian} for a sketch.

To track the boundary time evolution in the CFT Hilbert space {associated with the radial quantization}, it is more convenient to perform $\tau \rightarrow it$ in Eq.\eqref{eq:Emetric_global} and \eqref{eq:Ebrane_global}. The metric becomes
\begin{align}\label{eq:metric_global}
    ds^2 = d\rho^2 - \cosh^2 \rho~ dt^2 + \sinh^2 \rho ~d\phi^2,
\end{align}
with identification $(t,\phi) \sim (-t,\phi+\pi)$, and the brane profile is 
\begin{align}
    T^2 = \frac{\cosh^2 \rho}{\sinh^2 \rho - \tan^2 t}.
\end{align}
Note that the $(\eta,\theta_L, \phi)$ coordinate only covers the $t\in(-\pi/2, \pi/2)$ region while the $(\rho,t, \phi)$ coordinate can be extended to $t\in(-\infty, \infty)$ as shown in the right panel of Fig.\ref{fig:RP2_holo_dual_Lorentzian}. To satisfy $(t,\phi) \sim (2n\pi-t,\phi+\pi)$ for any $n\in\mathbb{Z}$, it is necessary and sufficient to impose $t\sim t + 2\pi$. As a result, the maximally extended geometry can be regarded as a $\mathbb{Z}_2$ quotient of the global AdS$_3$ (but not its universal cover) with dS$_2$ end-of-the-world branes.

Considering the symmetries, let us focus on the $t\in[0,\pi/2]$ region in the following. It is straightforward to see that at $t= \pi/2$, the dS brane approaches the asymptotic boundary $\rho\rightarrow\infty$, as sketched in Fig.\ref{fig:RP2_holo_dual_Lorentzian}.

This behavior can be understood as a consequence of the unitary time evolution in the dual CFT Hilbert space defined on a circle with length $2\pi$ as follows. At $t=0$, the CFT state is given by the transition matrix 
\begin{align}
    \mathcal{T}_{t=0} \propto |C\rangle\langle0|, 
\end{align}
prepared by the Euclidean path integral over $\mathbb{RP}^2$. Since $\ket{C}$ satisfies the crosscap condition Eq.\eqref{eq:Xstate_condition}, the transition matrix $\mathcal{T}_{t=0}$ is invariant under the following three global symmetry generators: $L_0 -\bar{L}_{0}, L_1 + \bar{L}_{-1}, L_{-1} + \bar{L}_{1}$ which satisfy \begin{align}\label{eq:Xcapstate_symmetry}
    &[L_0 - \bar{L}_0, L_1 + \bar{L}_{-1}] = -(L_1+\bar{L}_{-1}), \nonumber\\
    &[L_0 - \bar{L}_0, L_{-1} + \bar{L}_{1}] = (L_{-1}+\bar{L}_{1}), \\ 
    &[L_1 + \bar{L}_{-1}, L_{-1} + \bar{L}_{1}] = 2(L_{0}+\bar{L}_{0}),  \nonumber
\end{align}
and hence form an $SL(2,\mathbb{R}) \cong SO(2,1)$. 
On the gravity side, the domain of dependence of the $t=0$ time slice has two boost symmetries and one rotation symmetry, which generate an $SO(2,1)$. On the other hand, the three translations (two spacelike and one timelike) are broken. This is also precisely the isometry group of dS$_2$. 

On the other hand, at $t=\pi/2$, the CFT state is given by 
\begin{align}\label{eq:TM2pi}
    \mathcal{T}_{t=\frac{\pi}{2}} \propto e^{-i\pi H/2}|C\rangle\langle0| e^{i\pi H/2} \propto |B\rangle\langle0|, 
\end{align}
where $H = L_0 + \bar{L}_0 - \frac{c}{12}$ is the CFT Hamiltonian, and $\ket{B} \equiv e^{-i\pi H/2}\ket{C}$ satisfies 
\begin{align}\label{eq:Boundary_state}
    \left(L_n - \bar{L}_{-n}\right) \ket{B} = 0, \qquad n\in \mathbb{Z}.
\end{align}
This can be shown by noticing that 
\begin{align}
    L_n H^k = (n+H)^k L_n, 
\end{align}
and hence
\begin{align}
    L_n e^{itH} = e^{it(n+H)} L_n.
\end{align}
Accordingly, 
\begin{align}
    &(L_n-\bar{L}_{-n})e^{-i\left(\frac{1}{2}+k\right)\pi H}\ket{C} = 0, \\
    &(L_n-(-1)^n\bar{L}_{-n})e^{-ik\pi H}\ket{C} = 0. 
\end{align}
Eq.\eqref{eq:Boundary_state} is nothing but the condition satisfied by conformal boundary states \cite{Ishibashi88,Cardy89}, which gurantees there is no energy flux across the boundary. Since a conformal boundary state $|B\rangle$ does not contain real space entanglement \cite{MRTW14}, the transition matrix $\mathcal{T}_{t=\frac{\pi}{2}}\propto|B\rangle\langle0|$ has vanishing entanglement pseudo entropy \cite{NTTTW20} and
is expected to be dual to a trivial spacetime. This expectation is indeed realized in the proposed gravity dual, where the entire $t=\pi/2$ time slice and its causal diamond are nothing but void. From the point of view of symmetries, $\ket{B}$ is invariant under the following three global symmetry generators 
$L_0 -\bar{L}_{0}, L_1 - \bar{L}_{-1}, L_{-1} - \bar{L}_{1},$
which again form an $SL(2,\mathbb{R}) \cong SO(2,1)$. This is, however, a different $SO(2,1)$ from the one appearing in the crosscap state Eq.\eqref{eq:Xcapstate_symmetry}. The void bulk time slice at $t=\pi/2$ clearly satisfies the two spatial translation symmetries and the rotation symmetry, which again forms an $SO(2,1)$. On the other hand, it is not invariant under the time translation and the two boosts. 
This matching between the CFT transition matrix $|B\rangle\langle0|$ and the void spacetime at $t=\pi/2$ provides another strong and independent consistency check of our AdS/XCFT construction with dS branes. 

As a summary, we have established a correspondence between a bouncing dS$_2$ brane in AdS$_3$/$\mathbb{Z}_2$ and the unitary time evolution of the crosscap state, post-selected on the CFT ground state. 
This suggests that XCFT may be used to describe dS gravity. 
{This suggests that XCFT$_2$ may be used to describe dS$_2$ gravity coupled to an external
environment via the so-called double holography, in a similar way that BCFT$_2$ can be used
to describe AdS$_2$ gravity coupled to a non-gravitational heat bath \cite{AMMZ19}.}
{As a hint from an alternative point of view, it is intriguing to note that the crosscap state $\ket{C}$ is profoundly related to the Type II$_1$ von Neumann algebra, which is believed to describe the observables on the static patch of the dS spacetime \cite{CLPW22}, in the following way. Consider a chain of $2N$ spins aligned in a circle, and then produce EPR pairs between antipodal points, one gets 
\begin{align}\label{eq:EAP}
    \ket{\Psi} = \frac{1}{2^{N/2}} \bigotimes_{n=1}^{N} \left( \ket{\uparrow}_{i}\otimes\ket{\uparrow}_{i+N} + \ket{\downarrow}_{i}\otimes\ket{\downarrow}_{i+N} \right). 
\end{align}
While it is hard to write down the crosscap state $\ket{C}$ explicitly in general, it is known that $\ket{\Psi}$ serves as a generalization of crosscap states in a class of integrable spin chains \cite{CK21}, in the sense that it satisfies the matching between the energy-momentum tensor at the antipodal points and preserves infinitely many conserved charges.\footnote{Another interesting property of $\ket{\Psi}$ is, when restricted to a subsystem containing the first $N$ spins, $\ket{\Psi}$ is indistinguishable from the canonical Gibbs state with infinite temperature in any 1D spin system \cite{CY24}.} We note that, at $N\rightarrow\infty$, $\ket{\Psi}$ can be regarded as the cyclic separating vector used to construct a Type II$_1$ von Neumann algebra in \cite{vonNeumann39,Powers67,AW68,Witten21}. The Type II$_1$ von Neumann algebra is supported on exactly a half of the whole system, which indeed corresponds to a static patch in our setup.} {More details and future questions related to this perspective will be discussed in section \ref{sec:conclusions}.}

\section{Holographic XCFT on \texorpdfstring{$\mathbb{K}^2$}{K2}}

\begin{figure}
    \centering
    \includegraphics[width=16.5cm]{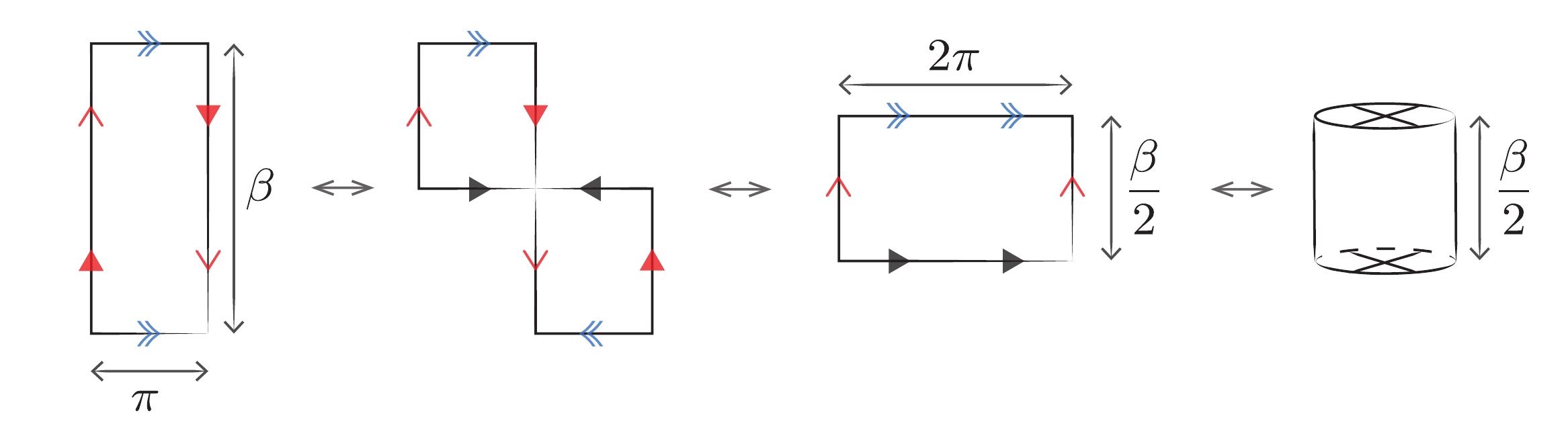}
    \caption{The standard definition of a Klein bottle (leftmost) and how it is related to $\langle C|e^{-\beta H/2}|C \rangle$ (rightmost) by sewing and gluing.}
    \label{fig:Klein_bottle}
\end{figure}

As a slightly more complicated example, let us consider a holographic XCFT defined on a Klein bottle $\mathbb{K}^2$. A standard definition of the Klein bottle and how it is equivalent to a cylinder with two crosscaps on the two ends is skeched in Fig.\ref{fig:Klein_bottle}. Let us consider the Klein bottle in Fig.\ref{fig:Klein_bottle}. Its partition function computes the inner product between two regulated crosscap states defined on the Hilbert space associated to a circle with length $2\pi$, 
\begin{align}
    Z_{\mathbb{K}^2} = \langle C|e^{-\beta H/2}|C \rangle.
\end{align}
It is also convenient to notice that such a Klein bottle can be obtained by starting from a torus $T^2$ parameterized by $\tau\in[-\beta/2,\beta/2)$, $\phi\in [0,2\pi)$ and then imposing the identification $(\tau,\phi)\sim(\beta/2-\tau,\phi+\pi)$. 
Let us restrict to the fundamental domain $\tau\in[-\beta/4,\beta/4]$. 

It is straightforward to find that there exist two bulk solutions. One has the metric 
\begin{align}
    ds^2 = d\rho^2 + \cosh^2 \rho~ d\tau^2 + \sinh^2 \rho ~d\phi^2,
\end{align}
and two disconnected pieces of branes located at  
\begin{align}
    T^2 = \frac{\cosh^2 \rho}{\sinh^2 \rho + \tanh^2 \left(\tau \pm \frac{\beta}{4} \right)}, 
\end{align}
respectively, with the identification $\phi\sim\phi+\pi$ at $\tau=\pm\frac{\beta}{4}$.
We call this the brane configuration. The bulk action $I^{\rm (brane)}$ computed from the brane configuration is 
\begin{align}\label{eq:brane_PF}
    I^{\rm (brane)} = -\frac{\beta}{16G_N} + \frac{{\rm arccoth}(-T)}{2G_N} = -\frac{c\beta}{24} - 2\log Z_{\mathbb{RP}^2} = -\frac{c\beta}{24} - 2S_{\rm xc},
\end{align}
where we have used Eq.\eqref{eq:pfunction}\eqref{eq:Sxc}. 
The second term is due to the existence of the two disconnected pieces of the end-of-the-world branes. 

Another solution has the metric 
\begin{align}
    ds^2 = d\rho^2 + \cosh^2 \rho~ \left(\frac{2\pi}{\beta}\right)^2 d\phi^2 + \sinh^2 \rho \left(\frac{2\pi}{\beta}\right)^2 ~d\tau^2,
\end{align}
with the identification $\phi\sim\phi+\pi$ at $\tau=\pm\frac{\beta}{4}$ and no branes. There are no singularities in the bulk. This is the well-known Euclidean BTZ geon geometry. The bulk action $I^{\rm (geon)}$ computed from the geon configuration is 
\begin{align}\label{eq:geon_PF}
    I^{\rm (geon)} = -\frac{\pi^2}{4G_N \beta} = -\frac{\pi^2 c}{6 \beta}. 
\end{align}
Note that $S_{\rm xc} = \log Z_{\mathbb{RP}^2}$ does not appear since there are no branes in the bulk. 

The holographic $\mathbb{K}^2$ partition function is given by 
\begin{align}
    Z_{\mathbb{K}^2} = \exp\left(-\min\{I^{\rm (brane)},I^{\rm (geon)}\}\right), 
\end{align}
and shows a first order phase transition at 
\begin{align}
    \beta = \sqrt{\left(\frac{24}{c}S_{\rm xc}\right)^2 + 4\pi^2} - \frac{24}{c} S_{\rm xc} \equiv  \beta_{*}.
\end{align}
The high temperature phase at $\beta<\beta_*$ is dominated by the geon geometry, and the low temperature phase at $\beta>\beta_*$ is dominated by the brane geometry. 

Similar to the Cardy formula for the torus partition function, the Klein bottle partition function has a universal behavior at the high temperature limit $\beta\rightarrow0$ and the low temperature limit $\beta\rightarrow\infty$ \cite{MR16,Tu17,Tsiares20}. The holographic partition functions computed from Eq.\eqref{eq:brane_PF} and \eqref{eq:geon_PF} precisely reproduce the universal behavior. While the $\beta$-dependent part is easy to recover just by taking a $\mathbb{Z}_2$ quotient of the thermal AdS geometry and the BTZ geometry, the end-of-the-world brane $Q$ plays an essential role in reproducing the $Z_{\mathbb{RP}^2}$-dependent part.

\section{Holographic \texorpdfstring{$p$-theorem}{p-theorem}}

So far we have established a correspondence between XCFT$_2$ and AdS$_3$ with dS$_2$ end-of-the-world branes, and checked that holographic computations match universal results in CFT. Let us then see an application of this AdS/XCFT correspondence. 

In 2D CFT, the central charge $c$ plays a central role and characterizes the degrees of freedom. It is known as the $c$-theorem that there exists a function (often called a $c$-function) which monotonically decreases under the RG flow and coincides the central charge $c$ at the critical points \cite{Zamolodchikov86}. Similarly, in 2D BCFT, the disk partition function (also called the $g$-function) characterizes the degrees of freedom of the boundary \cite{AL91} and monotonically decreases under the boundary RG flow, known as the $g$-theorem \cite{FK03}.\footnote{On the other hand, the $g$-function is not monotonic under the RG flow acting on the whole 2D ``bulk" \cite{GMS07}.} Given the similarity between the XCFT and the BCFT, it is natural to conjecture that there exists a $p$-function which monotonically decreases under some RG flow and coincides Eq.\eqref{eq:ZRP} at critical points. However, there is no proof of {such a} ``$p$-conjecture" so far. In fact, different from the boundary, since the crosscap is not a local concept, it is not even clear {how to implement an analogue of the boundary RG flow in the XCFT case.}\footnote{Similar to the BCFT case, the $p$-function in the XCFT is not monotonic under the RG flow which acts on the whole 2D ``bulk" \cite{TXWT18,CK21} in general, while it might be monotonic under certain restrictions \cite{CK21,ZHZTWT22}. }
Here, we will show that, in AdS/XCFT, there is a natural choice of the holographic RG flow. 
We prove a holographic $p$-theorem, following a similar strategy to the proof of the holographic $g$-theorem \cite{Takayanagi11,FTT11}. We expect that this holographic $p$-theorem should provide valuable hints to the $p$-conjecture on the XCFT side. 

Let us start by considering a holographic XCFT on $\mathbb{RP}^2$. Instead of its Euclidean gravity dual shown in Fig.\ref{fig:RP2_holo_dual}, let us consider its Lorentzian continuation shown in Fig.\ref{fig:RP2_holo_dual_Lorentzian}. It turns out to be useful to consider a Poincar\'e coordinate $(t_P,x_P,z_P)$ whose $t_P=0$ time slice is identified with the global $t=\frac{\pi}{2}$ time slice. Performing the coordinate transformation 
\begin{align}
    &\sqrt{1+\sinh^2\rho}~\cos \left(t-\frac{\pi}{2}\right) = \frac{z_P}{2} \left(1+\frac{1-t_P^2+x_P^2}{z_P^2}\right), \\
    &\sqrt{1+\sinh^2\rho}~\sin \left(t-\frac{\pi}{2}\right) = \frac{t_P}{z_P}, 
\end{align}
we get the standard Poincar\'e metric 
\begin{align}
    ds^2 = \frac{(dz_P)^2 - (dt_P)^2 + (dx_P)^2}{(z_P)^2}.
\end{align}
Let us again focus on the $0<t\leq\pi/2$ region, which is partially covered by $t_P\leq0$. The brane profile in the Poincar\'e coordinate is given by 
\begin{align}
    t_P = -\sqrt{\frac{T^2}{T^2-1}} z_P. 
\end{align}

Let us consider the holographic RG flow on the $z_P$ direction.
To probe this RG flow without changing the bulk AdS scale (and hence without changing the central charge $c$), let us put a brane-localized matter field on the $Q$ and consider the case where the matter configuration is translation invariant on the $x_P$ direction. In order for the brane-localized matter field to be senseible, we impose the null energy condition (NEC) on it: 
\begin{align}\label{eq:NEC}
    T_{ab}^Q u^a u^b \geq 0, 
\end{align}
where $T_{ab}^Q$ is the energy-stress tensor on the brane and $u^a$ can be any null vector {on the brane.} Due to the back reaction, the brane now has a new profile and let us denote it as 
\begin{align}
    t_P = f(z_P). 
\end{align}
We propose a $p$-function as 
\begin{align}
    p(z_P) = \exp\left(-\frac{{\rm arccoth}\left(-\frac{f'(z_P)}{\sqrt{\left(f'(z_P)\right)^2-1}}\right)}{4G_N}\right).
\end{align}
First of all, this matches Eq.\eqref{eq:pfunction} when there is no brane-localized matter and the bulk gravity describes a holographic XCFT. Secondly, if we choose $(u^t,u^x,u^z) = (f',\sqrt{(f')^2-1},1)$, then one can derive that $f''(z_P)\leq0$ from Eq.\eqref{eq:NEC}, and this further leads to $p'(z_P)\leq0$. Another quicker but less straightforward way to see this is by noticing that when the brane-localized matter satisfies NEC, two points on the brane that are connected by a casual curve on the brane should be always connected by a causal geodesic in the bulk \cite{Ishihara00,OW21}. Then $f''(z_P)\leq0$ must be satisfied otherwise one would find counterexamples to the statement above. 

As a summary, we have constructed $p(z_P)$ which monotonically decreases under the holographic RG flow on the $z_P$ direction. Based on the arguments above, we have derived a holographic $p$-theorem {without changing the $c$-function of the parent CFT.}

\section{Conclusions and Discussions}\label{sec:conclusions}

We have presented a minimal construction of gravity duals for 2D XCFTs, which is AdS$_3$ with dS$_2$ end-of-the-world branes.\footnote{The bulk geometries we constructed inherit the non-orientability from the boundary. It has been recently argued that non-orientable bulk saddle points should be taken into account even when the boundary CFT is orientable \cite{Yan22,Yan23,HN23}.} The construction is free from singularities thanks to the branes, and has a natural interpretation from cobordism theory point of view.
We first investigated {the holographic dual of }2D XCFT on $\mathbb{RP}^2$, computed the partition function (or equivalently $p$-function or crosscap entropy) and 1-point functions, saw the relation between the boundary state and the crosscap state $\ket{B}=e^{-i\frac{\pi}{2}H}\ket{C}$ naturally appears from the Wick rotation to the Lorentzian signature, and investigated {the holographic dual of }2D XCFT on $\mathbb{K}^2$.
\footnote{It is worth commenting that QFTs that do not admit a parity symmetry, e.g. chiral CFTs, cannot be consistently defined on nonorientable manifolds. It would be interesting to figure out how this manifests in the gravity dual of such theories, e.g. 3D chiral gravity discussed in \cite{LSS08}.}
We have also proven a holographic $p$-theorem as an application of our AdS/XCFT construction. From an alternative point of view, the searching of CFT dual to AdS$_3$ with negative tension dS$_2$ branes was initiated in \cite{AKTW20} and further explored in \cite{AKRTW21,AKRTW22,KRST23}. 
It was proposed in \cite{AKTW20,AKRTW21,AKRTW22,KRST23} that the dual should be a Lorentzian CFT with spacelike boundaries which, however, suffers from a divergence due to the non-normalizability of the boundary state. 
We conclude that a natural dual is a 2D XCFT, where a $\mathbb{Z}_2$ quotient is taken compared to the setup in \cite{AKTW20,AKRTW21,AKRTW22,KRST23}, and it is indeed related to the boundary state via $\ket{B}=e^{-i\frac{\pi}{2}H}\ket{C}$. 

{In the following, we would like to zoom in on some features of this construction and discuss related future directions.} 

\paragraph{dS holography via double holography}~\par
{An} interesting point we have made is that a dS$_2$ structure naturally appears in the bulk AdS$_3$, which is a warped product between dS$_2$ and a half-infinite line. It would be interesting to explore the possibility of using AdS/XCFT to describe dS holography, in the light of double holography \cite{AMMZ19}. We would like to emphasize that the UV-complete description here is the XCFT, which is local and unitary, and has a standard Hilbert space and time direction by construction. At least one of these features is often missed in dS holography proposals \cite{Strominger01,Witten01,Maldacena02,AKST04,AHS11,DST18,Susskind21,Susskind21-2,HNTT21}.
One may worry that, different from the standard double holography with positive tension AdS branes, there is no locally localized graviton on our negative tension dS branes via the standard Karch-Randall mechanism \cite{KR00,RS99-1,RS99-2} in the higher dimensional analogy of our setup. This problem may be got around by introducing a holographic CFT on the end-of-the-world brane, in a similar manner to \cite{Maldacena10}. 

\paragraph{Crosscap states and Type II$_1$ von Neumann algebra}~\par

{At the end of section \ref{sec:Lorentzian}, we briefly commented that the conformal crosscap state $\ket{C}$ can be naturally related to a Type II$_1$ von Neumann algebra, which is believed to describe observables in a static patch of dS \cite{CLPW22}. Let us make the statement more precise and discuss some related future directions. 

Starting from a crosscap state defined on a circle, we divide it into two semicircle subsystems. There are two distinct ways to relate the physics on one semicircle to a Type II$_1$ von Neumann algebra. 

The first way is as follows. Consider a compactified boson CFT, whose central charge is $c=1$. At a certain compactified radius, the lattice counterpart is given by the XXX spin chain. Consider such a spin chain defined on a circle with $2N$ spins aligned in order $i=1,2,3,...2N$. In this case, the lattice counterpart of the crosscap state is given by \eqref{eq:EAP} \cite{CK21}. Such a state appears in a classical construction of the Type II$_1$ von Neumann algebra \cite{vonNeumann39,Powers67,AW68}. Starting from \eqref{eq:EAP}, take the $N\to\infty$ limit. In this case, the algebra of operators acting on a finite subset of the first $N$ spins, which forms a semicircle subsystem in our case, after taken the closure by including weak limits, forms a Type II$_1$ von Neumann algebra.
A review of this construction can be found in, e.g. \cite{Witten21,CLPW22}. 
In this construction, the state \eqref{eq:EAP} at $N\rightarrow\infty$ serves as a special state called the  cyclic separating vector. 

The second way is as follows. Consider the transition matrix $|C\rangle\langle0|$ obtained by combining the crosscap state and the CFT ground state. The gravity dual in the AdS/XCFT construction describes a dS$_2/\mathbb{Z}_2$ EOW brane floating in the AdS$_3$ bulk, as discussed in section \ref{sec:Lorentzian}. Let us consider the semicircle subsystem of it. Focusing on the $t=0$ time slice, the semicircle subsystem is expected to correspond to exactly half of the bulk. The intersection between the causal diamond of this half-bulk region and the end-of-the-world brane is nothing but a dS$_2$ static patch. Interestingly, another famous way to construct a Type II$_1$ algebra, proposed in \cite{CLPW22}, is to consider the semiclassical gravity coupled to an observer system in the dS static patch. This suggests there is a chance that the semiclassical physics associated with the bulk causal diamond described above (possibly coupled to additional degrees of freedom identified as an observer) gives a Type II$_1$ von Neumann algebra. It would be interesting to make this connection rigorous. 
}

\paragraph{Extending AdS/XCFT and non-orientable holography to higher dimensions}~\par

{Another promising future direction is the extension to higher dimensions. In this paper, we have focused on the holography for 2D CFT defined on surfaces with crosscaps, which is equivalent to non-orientable surfaces. Although the crosscap is a notion in 2D, but we may straightforwardly extend it to higher dimensional manifolds, by cutting a codimension-0 ball-shaped hole and identifying antipodal points. However, different from 2D, the equivalence between the existence of crosscaps and orientability does not hold in higher dimensions. For example, $\mathbb{RP}^2 \times S^1$ is a 3D nonorientable manifold, but it does not possess a crosscap defined in the way above. On the other hand, $\mathbb{RP}^3$ includes one crosscap, but is orientable. Based on this, there are at least two directions of extensions to higher dimensions. The first one is holography for higher dimensional CFTs with crosscaps. We expect the relationship between dS branes and such crosscap CFTs can be directly extended to higher dimensions. The second direction is holography for higher dimensional nonorientable CFTs, which is expected to be way much more complicated. }

\section*{Acknowledgements}
I would like to in particular thank Andy Strominger and Tadashi Takayanagi for extensive discussions and insightful comments. 
I am also grateful to 
Levent Alp\"oge, Stefano Antonini, Raphael Bousso, Xujia Chen, Katie Ebner-Landy, Matthew Heydeman, Daniel Jafferis, Yikun Jiang, Alex May, Jake McNamara, Hidetoshi Omiya, Haoyu Sun, Marija Toma\v{s}evi\'{c}, Herman Verlinde, Diandian Wang and Yasushi Yoneta for stimulating conversations.
Finally, I would like to thank Shan-Ming Ruan, Tadashi Takayanagi and Diandian Wang for helpful comments on a draft of this paper.
I am supported by the Society of Fellows at Harvard University. 

\appendix

\bibliographystyle{jhep}
\bibliography{AdS_XCFT}

\end{document}